\documentclass[12pt]{article}
\usepackage{amsfonts,amssymb,amsmath,amscd}
\usepackage{stmaryrd}
\usepackage[all,dvips,arc,curve,color,frame]{xy}
\usepackage{graphicx}

\usepackage{pictexwd,dcpic}
\usepackage{comment}
\usepackage{longtable}
\usepackage{dsfont}
\DeclareGraphicsExtensions{.pdf,.png,.jpg,.eps}

\usepackage{color}

\def \TITLE {Superconformal vertex algebras in four dimensions}
\newcommand{\beq}{\begin{equation}}
\newcommand{\eeq}{\end{equation}}
\newcommand{\beqa}{\begin{eqnarray}}
\newcommand{\eeqa}{\end{eqnarray}}
\newcommand{\beqs}{\begin{eqnarray*}}
\newcommand{\eeqs}{\end{eqnarray*}}
\newcommand{\nn}{\nonumber \\}
\newcommand{\bnn}{\\ \nonumber}
\newcommand{\nnb}{\nonumber \\}

\def \ravno {\hspace{-3pt}&\hspace{-3pt}=\hspace{-3pt}&\hspace{-3pt}}
\def \podr {&& \hspace{-15pt}}

\newcommand{\Section}[1]{%
 \refstepcounter{section}
 \section*{\large \arabic{section}. #1}%
 \addtocontents{toc}{\protect\vspace{-8pt}}
 \addtocontents{toc}{\contentsline {section}{\thesection.\hspace{6pt}{#1}}{\arabic{page}}}}
\newcommand{\Subsection}[1]{%
 \refstepcounter{subsection}
 \subsection*{\bf\normalsize \arabic{section}.\arabic{subsection}. #1}%
 \addtocontents{toc}{\protect\vspace{-10pt}}
 \addtocontents{toc}{\contentsline {section}{\rm\protect\hspace{16pt}\thesubsection.\hspace{3pt}{#1}}{\arabic{page}}}}
\newcommand{\ASection}[1]{%
 \refstepcounter{section}
 \section*{\large Appendix \Alph{section}. #1}%
 \addtocontents{toc}{\protect\vspace{-8pt}}
 \addtocontents{toc}{\contentsline {section}{Appendix \thesection.\hspace{6pt}{#1}}{\arabic{page}}}}
\newcommand{\ASubsection}[1]{%
 \refstepcounter{subsection}
 \subsection*{\bf\normalsize \Alph{section}.\arabic{subsection}. #1}%
 \addtocontents{toc}{\protect\vspace{-10pt}}
 \addtocontents{toc}{\contentsline {section}{\rm\protect\hspace{16pt}\thesubsection.\hspace{3pt}{#1}}{\arabic{page}}}}

\newcounter{Theorem}
\newcounter{Definition}
\newcounter{Remark}
\newcounter{Example}

\def \setcntrs {\setcounter{equation}{0}\setcounter{Theorem}{0}\setcounter{Definition}{0}\setcounter{Remark}{0}\setcounter{Example}{0}}

\renewcommand{\thesection}{\arabic{section}}
\renewcommand{\theequation}{\arabic{section}.\arabic{equation}}
\renewcommand{\theTheorem}{\arabic{section}.\arabic{Theorem}}
\renewcommand{\theRemark}{\arabic{section}.\arabic{Remark}}
\renewcommand{\theDefinition}{\arabic{section}.\arabic{Definition}}
\renewcommand{\theExample}{\arabic{section}.\arabic{Example}}

\newenvironment{Proposition}[1][\bf Proposition \theTheorem]{%
        \refstepcounter{Theorem}\noindent\textbf{#1.}${}$\hspace{1pt}${}$\it}{}

\newcounter{tmpc}
\newlength{\tmplenght}
\newlength{\tmplenghta}
\newlength{\tmplenghtb}
\newlength{\tmplenghtc}

\newenvironment{LIST}[1]{%
\setlength{\tmplenghta}{#1}
\setlength{\tmplenghtb}{#1}
\setlength{\tmplenghtc}{#1}
\advance\tmplenghtb-5pt
\advance\tmplenghtc 42pt
\setcounter{tmpc}{0}
\begin{list}{{\rm (\alph{tmpc})}}{\usecounter{tmpc}
\setlength{\leftmargin}{\tmplenghta}
\setlength{\rightmargin}{0cm}
\setlength{\itemsep}{1pt}
\setlength{\topsep}{3pt}
\setlength{\labelsep}{5pt}
\setlength{\labelwidth}{\tmplenghtb}
\setlength{\listparindent}{\tmplenghta}}
}{\end{list}}

\newcommand{\Mbf}[1]{\ensuremath{\mathchoice
                    {\mbox{\boldmath$\displaystyle\mathbf{#1}$}}
                    {\mbox{\boldmath$\textstyle\mathbf{#1}$}}
                    {\mbox{\boldmath$\scriptstyle\mathbf{#1}$}}
                    {\mbox{\boldmath$\scriptscriptstyle\mathbf{#1}$}}}}

\DeclareMathAlphabet{\mathbbm}{U}{bbm}{m}{n}

\DeclareSymbolFont{ltrs}     {OT1}{pzc}{m}{it}
\DeclareSymbolFont{ltrsa}     {OMS}{cmsy}{m}{n}
\DeclareSymbolFont{ltrsA}{U}{txmia}{m}{it}
\DeclareSymbolFont{symbolsC}{U}{txsyc}{m}{n}
\DeclareSymbolFont{ltrsB}{U}{rsfs}{m}{n}

\DeclareSymbolFontAlphabet{\mfrak}{ltrsA}
\DeclareMathAlphabet{\mathpzc}{OT1}{pzc}{m}{it}
\DeclareMathAlphabet{\mathrsfs}{U}{rsfs}{m}{n}

\def\C{\mathbb{C}}
\def\R{\mathbb{R}}

\def\Z{\mathbb{Z}}
\def\M{\overline{M}}
\def\Hom{\text{\rm Hom}}

\def\Span{\text{\rm Span}}

\def\su{\text{\it su}}

\def\Spin{\text{\it Spin}}

\def\di{\partial} 
\def\DI{\Mbf{\di}}

\def\spr{\cdot}
\def\Spr{\hspace{1pt}\spr\hspace{1pt}}

\def\x{\mathrm{x}} 

\def\z{\mathrm{z}}

\def\a{\mathrm{a}}

\newcommand{\ovline}[1]{\bar{#1}}

\def\DIM{D}

\def\RSYM{\mathcal{A}}
\def\RCHA{\mathcal{R}}
\def\ETR{T}
\def\eTR{\text{\rm T}}

\newcommand{\EROT}[2]{\Omega_{#1\hspace{0.3pt},\hspace{1.2pt}#2}}
\newcommand{\DEL}[2]{\delta_{#1\hspace{0.3pt},\hspace{1.2pt}#2}}

\def\EDL{H}

\def\ESC{C}

\def\Bt{\text{\rm T}\hspace{-5.5pt}/\hspace{1pt}}
\def\BT{\text{\rm T}\hspace{-7.5pt}/\hspace{1pt}}
\def\CDF{\mathrsfs{D}}
\def\ADF{\mathrsfs{B}}
\def\CBDF{\ovline{\mathrsfs{D}}}
\def\ABDF{\ovline{\mathrsfs{B}}}
\def\EVO{Y}

\def\vac{|0\rangle}
\newcommand{\Prt}[1]{p_{#1}}

\def\alp{\alpha}
\def\balp{\dot{\alpha}}
\def\bet{\beta}
\def\bbet{\dot{\bet}}

\def\Ia{A}
\def\Ib{B}
\def\Ic{C}

\def\ETH{\Mbf{\theta}}
\def\EBTH{\ovline{\ETH}}

\newcommand{\Eth}[2]{\theta^{#1}_{#2}}   

\newcommand{\Ebth}[2]{\ovline{\theta}_{#1}^{#2}}   

\def\Est{Q}
\def\eST{\text{\rm Q}}
\def\eBST{\ovline{\text{\rm Q}}}

\newcommand{\EST}[2]{Q_{#1}^{#2}}   

\newcommand{\EBST}[2]{\ovline{Q}^{#1}_{#2}}   

\newcommand{\ESSC}[2]{S^{#1}_{#2}}   

\def\Essc{S}

\newcommand{\EBSS}[2]{\ovline{S}_{#1}^{#2}}   

\def\esig{\text{\small $\Sigma$}}

\newcommand{\Esig}[3]{(\esig^{#1}){}_{#2}^{#3}}   

\newcommand{\EsigD}[3]{(\esig_{#1}){}_{#2}^{#3}} 

\newcommand{\EDsig}[3]{(\ovline{\esig}{}^{#1}){}^{#2}_{#3}}   

\newcommand{\EDsigD}[3]{(\ovline{\esig}{}_{#1}){}^{#2}_{#3}} 

\newcommand{\ERsig}[4]{(\esig_{#1,#2}){}^{#3}_{#4}}   

\newcommand{\ERsigU}[4]{(\esig^{#1,#2}){}^{#3}_{#4}}   

\newcommand{\ERDsig}[4]{(\ovline{\esig}_{#1,#2}){}^{#3}_{#4}}   

\newcommand{\ERDsigU}[4]{(\ovline{\esig}^{#1,#2}){}^{#3}_{#4}} 

\newcommand{\LIECYCL}[4]{\mathcal{M}(#1;#2,#3,#4)}
\newcommand{\LIECYCLsh}[1]{\mathcal{M}(#1)}
\newcommand{\TLIEVEC}[4]{\widetilde{\mathcal{Z}}(#1;#2,#3,#4)}
\newcommand{\LIEVEC}[4]{\mathcal{Z}(#1;#2,#3,#4)}
\newcommand{\LIEVECsh}[1]{\mathcal{Z}(#1)}

\def\ESPIN{\mathrsfs{S}_E}
\def\EBSPIN{\ovline{\mathrsfs{S}}\hspace{-2.5pt}_E}

\newcommand{\vrestr}[2]{\!\left.\raisebox{#1}{$\,$}\!\right|_{\,
\raisebox{1pt}{\small \(#2\)}}}

\def\KSIP{}
\def\KSIM{-}

\def\Ksip{+}
\def\Ksim{-}

\allowdisplaybreaks[1]

\title{Superconformal vertex algebras\\in four dimensions}

\author{Dimitar Nedanovski\thanks{dnedanovski@inrne.bas.bg}}

\date{}

\begin{document}

\maketitle

\thispagestyle{empty}

\vspace{-0.5cm}

{\footnotesize
\begin{center}
University of Sofia, Physics Department,\\
5 James Bourchier Blvd., BG--1164 Sofia, Bulgaria
\\[0.4em]
Institute for Nuclear Research and Nuclear Energy,\\
72 Tsarigradsko Chaussee Blvd., BG--1784 Sofia, Bulgaria
\end{center}
}
\begin{center}
\date{\today}
\end{center}

\vspace{0cm}

\begin{abstract}
A superfield formalism for quantum fields with N-extended superconformal symmetry is developed using vertex algebra techniques in four dimensions.

\medskip

\noindent
Mathematics Subject Classification (2010): 17B69, 81T60
\end{abstract}%

\tableofcontents

\Section{Introduction}\label{Se1}
\setcntrs

A vertex algebra is an algebraic structure on the vector space of all local quantum fields\footnote{this space is also called the Borchers' class} in a given Quantum Field Theory (QFT), which reflects the concept of Operator Product Expansion (OPE),introduced by K. Wilson (\cite{W69}) as a tool for studying the short distance behavior in QFT.
There have been various further works (e.g., \cite{WZ72}, \cite{M77}) on how the OPE can be rigorously implemented in the general QFT framework or, derived from it.
However, up to now only in conformal QFT, and especially in the two--dimensional conformal filed theory, the notion of OPE has been fully understood at the axiomatic level and this has led to the concept of vertex algebra first introduced by Borcherds (\cite{Bor86}).

In higher spacetime dimensions the notion of a vertex algebra was generalized in \cite{N05} in a one--to--one correspondence with models of Wightman fields obeying the so called Global Conformal Invariance (GCI) as introduced in \cite{NT01}. 
GCI is an invariance (or, equivariance) of the Wightman functions under finite transformations of the two--fold spin covering of the geometric conformal group.

The present work extends the vertex algebra techniques to superconformal field theories, using the ideas of \cite{N05}.

\medskip

\noindent
{\bf Basic notations.}
We work in four dimensions, but most of the constructions we use are valid in arbitrary dimension and for them the definitions are given for general dimension $\DIM$.
Vectors in the $\DIM$-dimensional Minkowski space will be denoted by $\x=\bigl(x^{\mu}\bigr)_{\mu \, = \, 0}^{\DIM-1}\,$, $\x_j$ ($j=1,2,\dots$). We shall use also vectors in the complexified Euclidean space denoted by $\z=\bigr(z^{\mu}\bigr)_{\mu \, = \, 0}^{\DIM-1}\,$ (note that the labelling of coordinates starts from 0),  $\z_j$ ($j=1,2,\dots$), etc. .
The corresponding metrics (and scalar products) are,
$\x^2$ $\equiv$ $\x \spr \x$ $\equiv$ $-(x^0)^2$ $+$ $(x^1)^2$ $+$ $\cdots$ $+$ $(x^{\DIM-1})^2$
and
$\z^2$ $\equiv$ $\z \spr \z$ $\equiv$ $(z^0)^2$ $+$ $\cdots$ $+$ $(z^{\DIM-1})^2$.
Derivatives like $\frac{\di}{\di z^{\mu}}$ will be shorten as $\di_{z^{\mu}}$. 
Throughout this paper the Einstein convention about summation over repeated indices is assumed. 
\Section{Vertex algebras: a synopsis}\label{Se2}\label{Se-VA}
\setcntrs

In this section we make a short review of vertex algebras and conformal invariance.
We shall follow with minor modifications the notations of \cite{BN06}, \cite{N05}, \cite{NT05}. 

GCI allows us to extend the QFT models with such an invariance on a compactification of the real spacetime. 
The latter is the (conformally) compactified Minkowski space $\M$.
There are special complex coordinates that are obtained by a complex conformal transformation, which globally cover $\M$. Vertex operators correspond to local quantum fields in these new coordinates.
This formalism is known in the literature as \emph{compact picture},
because of the compactness of the real spacetime in this representation.

Let us stress two important technical features related to the formalism of vertex algebras in GCI QFT.

The first is that in the complex coordinates that parametrize 
$\M$
it is natural to change the signature of the metric to a complexified Euclidean metric.
In this way the vertex operator depends on a formal complex Euclidean variable $\z = (z^0,\dots,z^{\DIM-1})$ $\in$ $\C^{\DIM}$.
Furthermore, the natural generators of the conformal symmetry in these coordinates form a real basis of the Euclidean conformal Lie algebra.
However, this {\bf \emph{does not mean}} that we are considering Euclidean fields in the sense of Euclidean field theory!
The point is that the relevant real structure in the compact picture, which comes from the initial theory on the Minkowski space is not an ordinary complex conjugation related to the new coordinates or the symmetry generators.
The ``physical'' real structure in described in Appendix \ref{SeZ4}.

Second,
the vertex operators are not exactly quantum fields in the usual sense of Wightman axioms as they are not distributions. 
They are considered as formal power series in the spatial coordinates (the above complex coordinates).
This is just for convenience and it can be considered as a topological lighten of the formalism: instead of with actual distributions we work with formal distributions (as these formal series are called in \cite{K98}).
However, the axioms of vertex algebras are strong enough to allow us to prove that the vertex operators are not only formal distributions but determine also actual~distributions.

The theory of vertex algebras is based on the formalism of formal Laurent--Taylor series with ``light--cone poles''.
This formalism can be found in \cite[Sect. 1]{N05} or in \cite[Sect I]{BN06}.
However, in the present work these techniques will not play a central role and so, we shall not review them.

\medskip

\noindent \textbf{Vertex algebras}

\medskip

We shall mainly follow the definition of a vertex algebra as given in \cite[Definition 2.1]{N05}, but for the sake of completeness we give the full definition below.
For a short review of the definitions and especially for a comparison with the one--dimensional, chiral case, we refer the reader to Sects. 1 and 2 of \cite{BN08}.
\medskip

Thus, a \textit{vertex algebra} is a $(\Z/2\Z)$--graded vector space $V$ endowed with an even (i.e., parity preserving) (bi)linear map\footnote{It defines a homomorphism
$V\longrightarrow End(V [[\z]][(\z^2)^{-1}])$ called \textit{ vertex operator}. Sometimes $\EVO (a,\z) \, b$ is also denoted by $a(\z)b$.}\footnote{Let us remark that instead of defining individually the vertex operators $\EVO (a,\z)$ like two sided infinite formal series as it was done in \cite[Definition 2.1]{N05} one can use, without any loss of generality, the approach of \cite[Sect. 1]{BN08}.
In this way, $\EVO (a,\z)$ can be defined not individually but only when applied on $b \in V$ and then the result lies in the space of Laurent--Taylor formal series $V [[\z]][(\z^2)^{-1}]$.
Equalities like 
$\mathop{\sum}\limits_{k \, = \, 1}^m \Gamma_k \EVO (\a_k, \z) = 0$
where $\Gamma_k$ are differential operators just mean that  
$\mathop{\sum}\limits_{k \, = \, 1}^m \Gamma_k \EVO (\a_k, \z) \, b= 0$
for all $b \in V$.}
$$
V \otimes V \, \ni \, a \otimes b \, \mapsto \, \EVO (a,\z) \, b \, \in V [[\z]][(\z^2)^{-1}]\,,
$$
($V[[\z]][(\z^2)^{-1}]$ stands for the space of formal Laurent--Taylor series with poles at $\z^2=0$, $\z\in\C^\DIM$), a set of mutually commuting even endomorphisms $\ETR_{\mu}$ for $\mu =0,\dots,\DIM-1$
called \textit{(infinitesimal) translation endomorphisms}, and
an even vector $\vac \in V$ called a \textit{vacuum}.
These data are subject to the following axioms:
\begin{LIST}{33pt}
\item[$(a)$]
{\it Locality.} For all $a_1,a_2,b \in V$ we have
\beqa\label{LocAx}
\podr
\bigl((\z_1-\z_2)^2\bigr)^{N_{a_1,a_2}} \, \EVO(a_1,\z_1) \, \EVO(a_2,\z_2) \, b 
\nnb \podr = \,
(-1)^{\Prt{a_1}\Prt{a_2}}
\bigl((\z_1-\z_2)^2\bigr)^{N_{a_1,a_2}} \, \EVO(a_2,\z_2) \, \EVO(a_1,\z_1) \, b \,,
\eeqa
where $\Prt{a_j}$ is the parity of $a_j$ and $N_{a_1,a_2}$ is a nonnegative integer;
another way of writing the above equality is as vanishing of the super-commutator
$[\EVO(a_1,\z_1),\EVO(a_2,\z_2)]$ when multiplied with a sufficiently large integral power of $(\z_1-\z_2)^2$. 
\item[$(b)$]
{\it Translation covariance.}
For all $a \in V$ and $\mu = 0,\dots,\DIM-1$ we have
\beq\label{TrInvAx}
[\ETR_{\mu},\EVO (a,\z)] \, b \, := \, \ETR_{\mu} \, (\EVO (a,\z) \, b) - \EVO (a,\z) \hspace{1pt} (\ETR_{\mu} \, b) \, = \, \di_{z^{\mu}} \, (\EVO (a,\z) \, b).
\eeq
(This again can be written shortly as 
$[\ETR_{\mu},\EVO (a,\z)] = \di_{z^{\mu}} \, \EVO (a,\z)$.)
\item[$(c)$]
{\it Vacuum axiom.}
We have $\ETR_{\mu} \vac = 0$ for every $\mu = 0,\dots,\DIM-1$ and for every $a \in V$:
\beq\label{VacAx}
\EVO(\vac,\z) \, a \, = \, a\,,\quad
\EVO(a,\z) \, \vac \, \in \, V[[\z]]\,,\quad
\EVO(a,\z) \, \vac \vrestr{12pt}{\z \, = \, 0} \, = \, a \,.
\eeq
\end{LIST}

\medskip

This is the definition of a vertex algebra.
Before stating the main properties of vertex algebras which we shall use, let us make one more definition:
a \textit {local field} in a vertex algebra is a linear map
$$
a \mapsto \phi (\z) a \, \in \, V [[\z]][(\z^2)^{-1}]
$$
such that the super-commutator
$$
[\phi (\z_1),\EVO(b,\z_2)] \, a \, := \, 
\phi (\z_1)\,\EVO(b,\z_2) \, a - (-1)^{\Prt{\phi}\Prt{b}} \, \EVO(b,\z_2)\,\phi (\z_1) \, a
$$
is \emph{local} in the sense that it vanishes when multiplied with a sufficiently large power of $(\z_1-z_2)^2$ (\cite[Sect. IV.A]{BN06}):
\beq\label{EN2.4xx}
\bigl((\z_1-\z_2)^2)\bigr)^{N_{\phi,b}} \,
[\phi (\z_1),\EVO(b,\z_2)] \, a \, = \, 0 \,.
\eeq
The local field $\phi (\z)$ is additionally called {\it translation--invariant} if
$$
[\ETR_{\mu},\phi (\z)] \, a \, := \, \ETR_{\mu} \, (\phi (\z) \, a) - \phi (\z) \hspace{1pt} (\ETR_{\mu} \, a) \, = \, \di_{z^{\mu}} \, (\phi (\z) \, a)
$$
for all $\mu = 0,\dots,\DIM-1$ and $a \in V$.

According to \cite[Corollary 4.3]{BN06}, every translation--invariant local field $a \mapsto \phi (\z) \, a$ is of a form $a \mapsto \EVO(b,\z) \, a$ for some $b \in V$, i.e., it can be represented by a vertex operator.
In fact, $b = \phi (\z) \vac\vrestr{12pt}{\z \, = \, 0}$. 
The latter is an equivalent formulation of the Reeh--Schlieder theorem within the formalism of vertex algebras.

Translation--invariance of the vertex operators gives that (\cite[Proposition 3.2 (b)]{N05}):
\beq\label{YonVac}
\EVO (a,\z) \, \vac \, = \, e^{\z \Spr \eTR} \, a \,,
\eeq
where $\z \Spr \eTR := z^{\mu} \ETR_{\mu}$.

\medskip

\noindent\textbf{Vertex algebras and conformal symmetry}

\medskip

We say that the vertex algebra  $V$ is \textbf{conformal vertex algebra} (following \cite{N05}, with slight modifications) if we have an action of the conformal Lie algebra (see \eqref{ConCR}) on the space $V$ realized via even endomorphisms such that the vacuum is invariant under this action and
\beqa\label{CVA1}
\hspace{-35pt}
\bigl[H,\EVO(a,\z)\bigr]b
\, = \podr \EVO(H\hspace{1pt}a,\z)b + \z\hspace{-1pt}\cdot\hspace{-1pt}\di_\z \, \EVO(a,\z)b,
\\ \label{CVA2}
\hspace{-35pt}
\bigl[\EROT{\mu}{\nu},\EVO(a,\z)\bigr]b
\, =  \podr \EVO(\EROT{\mu}{\nu}\hspace{2pt}a,\z)b
+ (z^\mu \di_{z^\nu} - z^\nu \di_{z^\mu}) \, \EVO(a,\z)b,
\\ \label{CVA3}
\hspace{-35pt}
\bigl[\ESC_\mu,\EVO(a,\z)\bigr] b
\, = \podr \EVO(\ESC_\mu\hspace{1pt}a,\z)b - 2 \, z^\mu \, \EVO(H\hspace{1pt}a,\z)b
- 2 \, 
z^\nu \, \EVO(\EROT{\mu}{\nu}\hspace{2pt}a,\z)b
\nn \hspace{-35pt} \podr
+ \ (\z^2 \, \di_{z^\mu} - 2 \, z^\mu \, \z\hspace{-1pt}\cdot\hspace{-1pt}\di_\z) \, \EVO(a,\z)b.
\eeqa

\Section{Superconformal vertex algebras}\label{NSe3}
\setcntrs

\Subsection{Some preliminary notations}\label{Se2.1}
The Grassmann variables attached to the complexified four-dimensional Euclidean space are denoted by $\ETH$ $=$ $(\Eth{\alp}{\Ia})$ and $\EBTH$ $=$ $(\Ebth{\balp}{\Ia})$,where $\alp=1,2$, $\balp=\dot{1},\dot{2}$ are chiral spinorial indices and $A=1,\dots,N$ is an $su(N)$--index. Grassman variables with undoted spinorial indices are related to $(\frac{1}{2},0)$ and those with doted ones to $(0,\frac{1}{2})$ representations of the orthogonal Lie algebra.

Description of the superconformal Lie algebra is given in Appendix \ref{superconformal_Lie_algebra_definiton}. In this algebra with $\EST{\alp}{\Ia}$ and $\EBST{\balp}{\Ia}$ we denote the generators of \textit{supertranslations}. (Indices are the same as ones explained in the above paragraph.) The translation generators are denoted by $\ETR_\mu$, $\mu=0,1,2,3.$

We adopt the following conventions: $\ETH\spr\eST:=\Eth{\alp}{\Ia}\EST{\alp}{\Ia}$, 
$\EBTH\spr\eBST:=\Ebth{\balp}{\Ia}\EBST{\balp}{\Ia}$. For the \emph{left} derivatives $\mathop{\frac{\di}{\di \Eth{\alp}{\Ia}}}\limits^{\rightarrow}$ and $\mathop{\frac{\di}{\di \Ebth{\alp}{\Ia}}}\limits^{\rightarrow}$ we use the short notations $\di_{\Eth{\alp}{\Ia}}$ and $\di_{\Ebth{\balp}{\Ia}}$, respectively.

If $V=V_0 \oplus V_1$ is a 
$(\Z/2\Z)$--graded vector space,
then $V[\ETH,\EBTH]$ is the space of polynomials in the anti-commuting variables $(\ETH,\EBTH)$, which naturally is a vector super-space. Note that if, in addition, $V$ is a Lie super-algebra, then $V[\ETH,\EBTH]$ is again a Lie super-algebra.

\Subsection{Super vertex operators}

Let $\EVO (a,\z),$ $\z\in\C^4$ be a vertex operator from a vertex algebra endowed with an action of the superconformal algebra (defined by Proposition \ref{Pr3N1}).
We define \textit{super vertex operators} by the formula
\beq\label{EVO}
\EVO (a,\z,\ETH,\EBTH) \, := \, e^{\ETH \Spr \eST\,+\,
\EBTH \Spr \eBST
} \, \EVO (a,\z) \, e^{-\,\ETH \Spr \eST\,-\,
\EBTH \Spr \eBST
} \,.
\eeq
Note that $\EVO (a,\z,\ETH,\EBTH)$ is actually a polynomial in $\ETH,\EBTH$ with coefficients that are local translation--invariant fields.
Therefore, using the state--field correspondence
(\ref{YonVac}), $\EVO (a,\z,\ETH,\EBTH)$ can be reconstructed from its action on the vacuum,
\beqa\label{EVO-VAC}
\EVO (a,\z,\ETH,\EBTH) \, \vac \podr = \,
e^{\ETH \Spr \eST\,+\,
\EBTH \Spr \eBST
} \, \EVO (a,\z) \, \vac 
\, = \,
e^{\z \Spr \eTR\,+\,\ETH \Spr \eST\,+\,
\EBTH \Spr \eBST
} \, a.
\eeqa
This allows us to deduce the covariance properties of the so defined super vertex operators.
First, for the translations we have,
\beqa\label{T-VO-VAC}
\ETR_{\mu} \, \EVO (a,\z,\ETH,\EBTH) \, \vac \podr = \,
\ETR_{\mu} \, e^{\z \Spr \eTR\,+\,\ETH \Spr \eST\,+\,
\EBTH \Spr \eBST
} \, a \, = \,
e^{\z \Spr \eTR\,+\,\ETH \Spr \eST\,+\,
\EBTH \Spr \eBST
} \, \ETR_{\mu} \, a 
\nnb
\podr = \,
\EVO (\ETR_{\mu} \, a,\z,\ETH,\EBTH) \, \vac
\\ \label{T-VO-VAC2}
\podr = \, 
\di_{z^{\mu}} \, e^{\z \Spr \eTR\,+\,\ETH \Spr \eST\,+\,
\EBTH \Spr \eBST
} \, a 
\, = \,
\di_{z^{\mu}} \, \EVO (a,\z,\ETH,\EBTH) \, \vac
\qquad
\eeqa
and hence, by the Reeh--Schlieder theorrem we conclude:
\beq\label{T-VO}
\bigl[\ETR_{\mu} , \EVO (a,\z,\ETH,\EBTH) \bigr] \, = \,
\EVO (\ETR_{\mu} \, a,\z,\ETH,\EBTH)
\, = \,
\di_{z^{\mu}} \, \EVO (a,\z,\ETH,\EBTH) \,.
\eeq
Next, for the supertranslations we obtain
\beqa\label{Q-VO-VAC1}
\podr \hspace{-5pt}
\EST{\alp}{\Ia} \, \EVO (a,\z,\ETH,\EBTH) \, \vac =
\EST{\alp}{\Ia} \, e^{\z \Spr \eTR\,+\,\ETH \Spr \eST\,+\,
\EBTH \Spr \eBST
} \, a =
e^{\z \Spr \eTR} \, \EST{\alp}{\Ia} \, 
e^{
\KSIP
\ETH \spr \Bt \spr \EBTH} 
\, e^{\ETH \Spr \eST} \, e^{
\EBTH \Spr \eBST
} \, a 
\nnb
\hspace{-5pt}
\podr = \, 
\CDF_{\alp}^{\Ia} \, e^{\z \Spr \eTR} \, 
e^{
\KSIP
\ETH \spr \Bt \spr \EBTH} 
\, e^{\ETH \Spr \eST} \, e^{
\EBTH \Spr \eBST
} \, a \, = \,
\CDF_{\alp}^{\Ia} \, \EVO (a,\z,\ETH,\EBTH) \, \vac \,,
\eeqa
where
\beq\label{CDF}
\CDF_{\alp}^{\Ia} \, := \,
\di_{\Eth{\alp}{\Ia}} \, 
\Ksim \, 
\Esig{\mu}{\alp}{\bbet} \, \Ebth{\bbet}{\Ia} \, \di_{z^{\mu}} \,,
\eeq
since
\beq\label{QbQ}
\bigl[\ETH \spr \eST\hspace{1pt},\, 
\EBTH \spr \eBST 
\bigr] \, \equiv \,
-\,\Eth{\alp}{\Ia}\,\Ebth{\bbet}{\Ib} \bigl[ \EST{\alp}{\Ia}\hspace{1pt}, \, 
\EBST{\bbet}{\Ib}  
\bigr]
\, = \,
\KSIM\,
2 
\, \Eth{\alp}{\Ia} \, \Esig{\mu}{\alp}{\bbet} \, \Ebth{\bbet}{\Ia} \, \ETR_{\mu} \, =: \,
\KSIM\,
2 
\, \ETH \spr \BT \spr \EBTH \,.
\qquad
\eeq
Similarly, we compute
\beqa\label{Q-VO-VAC2}
\podr \hspace{-5pt}
\EVO (\EST{\alp}{\Ia} \, a,\z,\ETH,\EBTH) \, \vac =
e^{\z \Spr \eTR\,+\,\ETH \Spr \eST\,+\,
\EBTH \Spr \eBST
} \, \EST{\alp}{\Ia} \, a =
e^{\z \Spr \eTR} \, 
e^{
\KSIM
\ETH \spr \Bt \spr \EBTH} 
\, e^{
\EBTH \Spr \eBST
} \, e^{\ETH \Spr \eST} \, \EST{\alp}{\Ia} \, a 
\nnb
\hspace{-5pt}
\podr = \, 
\ADF_{\alp}^{\Ia} \, e^{\z \Spr \eTR} \, 
e^{
\KSIM
\ETH \spr \Bt \spr \EBTH} 
\, e^{
\EBTH \Spr \eBST
} \, e^{\ETH \Spr \eST} \, a \, = \,
\ADF_{\alp}^{\Ia} \, \EVO (a,\z,\ETH,\EBTH) \, \vac \,,
\eeqa
where
\beq\label{ADF}
\ADF_{\alp}^{\Ia} \, := \,
\di_{\Eth{\alp}{\Ia}} \, 
\Ksip \,
\Esig{\mu}{\alp}{\bbet} \, \Ebth{\bbet}{\Ia} \, \di_{z^{\mu}}
\eeq
and the same calculations for $\EBST{\balp}{\Ia}$ give
\beqa\label{BQ-VO-VAC1}
\EBST{\balp}{\Ia} \, \EVO (a,\z,\ETH,\EBTH) \, \vac \podr = \,
\CBDF^{\balp}_{\Ia} \, \EVO (a,\z,\ETH,\EBTH) \, \vac \,,
\\ \label{BQ-VO-VAC2}
\EVO (\EBST{\balp}{\Ia} \, a,\z,\ETH,\EBTH) \, \vac \podr = \,
\ABDF^{\balp}_{\Ia} \, \EVO (a,\z,\ETH,\EBTH) \, \vac\,,
\eeqa
with
\beq\label{CBDF}
\CBDF^{\bbet}_{\Ia} \, := \,
\di_{\Ebth{\bbet}{\Ia}} \, 
\Ksim \,
\Eth{\alp}{\Ia} \Esig{\mu}{\alp}{\bbet} \, \di_{z^{\mu}}\,,
\eeq
\beq\label{ABDF}
\ABDF^{\bbet}_{\Ia} \, := \,
\di_{\Ebth{\bbet}{\Ia}} \, 
\Ksip\,
\Eth{\alp}{\Ia} \Esig{\mu}{\alp}{\bbet} \, \di_{z^{\mu}}
\,.
\eeq
So, by the Reeh--Schlieder theorem we conclude again:
\beqa\label{Q-VO-1}
\bigl[\EST{\alp}{\Ia} \hspace{1pt},\, \EVO (a,\z,\ETH,\EBTH) \bigr] \podr = \,
\CDF_{\alp}^{\Ia} \, \EVO (a,\z,\ETH,\EBTH) \,,
\\ \label{Q-VO-2}
\EVO (\EST{\alp}{\Ia} \, a,\z,\ETH,\EBTH) \podr = \,
\ADF_{\alp}^{\Ia} \, \EVO (a,\z,\ETH,\EBTH) \,,
\\ \label{BQ-VO-1}
\bigl[\EBST{\balp}{\Ia} \hspace{1pt},\, \EVO (a,\z,\ETH,\EBTH) \bigr] \podr = \,
\CBDF^{\balp}_{\Ia} \, \EVO (a,\z,\ETH,\EBTH) \,,
\\ \label{BQ-VO-2}
\EVO (\EBST{\balp}{\Ia} \, a,\z,\ETH,\EBTH) \podr = \,
\ABDF^{\balp}_{\Ia} \, \EVO (a,\z,\ETH,\EBTH) \,.
\eeqa

Let $X$ be a generator of the supreconformal Lie algebra. Commutators 
$\bigl[X,\EVO(a,\z,\ETH,\EBTH)\bigr]$ are computed from their action on the vacuum
$\bigl[X,$ $\EVO(a,\z,\ETH,\EBTH)\bigr]\, \vac\,$ $=\,X \, \EVO(a,\z,\ETH,\EBTH) \, \vac$  and using the general formula \cite{BN}, \cite{M09}
\beq
 X \, \EVO(a,\z,\ETH,\EBTH) \, \vac \, = \,
\EVO (e^{-\text{\rm ad} (\z \Spr \eTR\,+\,\ETH \Spr \eST\,+\,
\EBTH \Spr \eBST
)} \bigl(X\bigr) \, a,\z,\ETH,\EBTH) \, \vac \,.
\eeq
Hence,
\beq\label{XCOM}
\bigl[ X \hspace{1pt},\, \EVO(a,\z,\ETH,\EBTH) \bigr] \, = \,
\EVO (e^{-\text{\rm ad} (\z \Spr \eTR\,+\,\ETH \Spr \eST\,+\,
\EBTH \Spr \eBST
)} \bigl(X\bigr) \, a,\z,\ETH,\EBTH) \,.
\eeq

Due to the nilpotency of $\text{\rm ad} (\z \Spr \eTR\,+\,\ETH \Spr \eST\,+\,
\EBTH \Spr \eBST
)$,
 $\, e^{-\text{\rm ad} (\z \Spr \eTR\,+\,\ETH \Spr \eST\,+\,
\EBTH \Spr \eBST
)} \bigl(X\bigr) $ is a polynomial in $\z,\ETH,\EBTH $ with coefficients in the superconformal Lie algebra and linearly depending of $X$. It can be splitted additively in two parts:
\beq
e^{-\text{\rm ad} (\z \Spr \eTR\,+\,\ETH \Spr \eST\,+\,
\EBTH \Spr \eBST
)} \bigl(X\bigr)
\,=\,
\LIECYCL{X}{\z}{\ETH}{\EBTH}
\,+\,
\TLIEVEC{X}{\z}{\ETH}{\EBTH},
\eeq
where $ \LIECYCL{X}{\z}{\ETH}{\EBTH} $ has coefficients belonging to 
$ \mathrm{Span}\{\ESC_{\mu},$ $\ESSC{\alp}{\Ia},$ $\EBSS{\balp}{\Ia},$ $\EROT{\mu}{\nu},$ $\EDL,$ 
 $\RSYM^{\Ia}_{\Ib},\RCHA\} $ and $ \TLIEVEC{X}{\z}{\ETH}{\EBTH} $ has coefficients belonging to $\mathrm{Span}\{\ETR_{\mu},$ $\EST{\alp}{\Ia},$ $\EBST{\balp}{\Ia}\} $. Using 
equations \eqref{T-VO}, \eqref{Q-VO-2} and \eqref{BQ-VO-2}, we can rewrite 
\beq
\EVO\bigl(\TLIEVEC{X}{\z}{\ETH}{\EBTH}\,a,z,\ETH,\EBTH\bigr)
\,=\,
\LIEVEC{X}{\z}{\ETH}{\EBTH}\EVO(a,\z,\ETH,\EBTH),
\eeq
where $\LIEVEC{X}{\z}{\ETH}{\EBTH}$ is first order differential operator (i.e., a vector field) in
$\z,\ETH,\EBTH$ with polynomial coefficients in $\z,\ETH,\EBTH$. Hence, the generic form of the commutators is
\begin{gather}
\label{EqY-Z-com}
\bigl[ X\hspace{1pt},\,\EVO(a,\z,\ETH,\EBTH)\bigr]
\,=\,
\LIEVEC{X}{\z}{\ETH}{\EBTH}\EVO(a,\z,\ETH,\EBTH)
\,+\,
\EVO\bigl(\LIECYCL{X}{\z}{\ETH}{\EBTH}\,a,\z,\ETH,\EBTH\bigr).
\end{gather}
The super-Jacobi identity for $\bigl[\,[X,\,X'],\,\EVO(a,\z,\ETH,\EBTH)\bigr]$ implicates:
\beqa\label{ZCOM1}
&&-\,\LIEVEC{[X,X']}{\z}{\ETH}{\EBTH}
\,+\,
\LIECYCL{[X,X']}{\z}{\ETH}{\EBTH}
\,=\,
\\
&&\bigl[-\,\LIEVEC{X}{\z}{\ETH}{\EBTH}\,+\,\LIECYCL{X}{\z}{\ETH}{\EBTH}\,,
\,-\,\LIEVEC{X'}{\z}{\ETH}{\EBTH}\,+\,\LIECYCL{X'}{\z}{\ETH}{\EBTH}
\bigr]\nonumber
\eeqa
(commutators are understood as $(\Z/2\Z)$--graded commutators).
Note that in Eq. (\ref{ZCOM1}) commutators like 
\(\bigl[
\LIEVEC{X}{\z}{\ETH}{\EBTH},
\LIECYCL{X'}{\z}{\ETH}{\EBTH}
\bigr]\)
are understood as a commutator of first and zeroth order differential operators in 
$(\z,\ETH,\EBTH)$.
Equation (\ref{ZCOM1}) is verified using the following results, in which shorten notations $\LIECYCL{X}{\z}{\ETH}{\EBTH}\equiv\LIECYCLsh{X}$ and $\LIEVEC{X}{\z}{\ETH}{\EBTH}$ $\equiv$ $\LIEVECsh{X}$ are used:
\beqa
\LIEVECsh{\ETR_\mu}
&=&
\di_{z^{\mu}},\quad
\LIEVECsh{\EST{\alp}{\Ia}}\,=\,\CDF_{\alp}^{\Ia},\quad
\LIEVECsh{\EBST{\balp}{\Ia}}\,=\,\CBDF^{\balp}_{\Ia}
\,,
\nonumber\\[0.75em]
\LIEVECsh{\EDL}
&=&
\z\spr\di_{\z}
\,+\,\frac{1}{2}\,\ETH\Spr\DI_{\ETH}
\,+\,
\frac{1}{2}\,\EBTH\Spr\bar{\DI}_{\EBTH}
\,,
\nn[0.75em]
\LIEVECsh{\EROT{\mu}{\nu}}
&=&
z_\mu\di_{z^{\nu}}-z_\nu\di_{z^{\mu}}
\,+\,
\Eth{\alp}{\Ia}\,\ERsig{\mu}{\nu}{\bet}{\alp}\,\di_{\Eth{\bet}{\Ia}}
\,-\,
\Ebth{\bbet}{\Ia}\,\ERDsig{\mu}{\nu}{\bbet}{\balp}\,\overline{\di}_{\Ebth{\balp}{\Ia}}
\,,
\nn[0.75em]
\LIEVECsh{\ESC_\mu}
&=&
\,-\,
2\,z_\mu\,\z.\,\di_{\z}
\,+\,
\z^2\,\di_{z^\mu}
\,+\,
2\,\Eth{\alp}{F}\,\Ebth{\bbet}{D}\,\EsigD{\mu}{\alp}{\bbet}\,\Eth{\gamma}{D}\,\Ebth{\dot{\sigma}}{F}\,\Esig{\nu}{\gamma}{\dot{\sigma}}\,\di_{z^\nu}
\nonumber\\*[0.2em]
& &
\,+\,
2\,z^{\nu}\,\Eth{\alp}{\Ia}\,\ERsig{\nu}{\mu}{\bet}{\alp}\,\di_{\Eth{\bet}{\Ia}}
\,-\,
z_\mu\,\ETH\cdot\DI_{\ETH}
\,\Ksip\,
2\,\Eth{\alp}{F}\,\Ebth{\bbet}{D}\,\EsigD{\mu}{\alp}{\bbet}\,\Eth{\gamma}{D}\,\di_{\Eth{\gamma}{F}}
\nonumber\\*[0.2em]
& &
\,-\,
2\,z^\nu\,\Ebth{\bbet}{\Ia}\,\ERDsig{\nu}{\mu}{\bbet}{\balp}\,\bar{\di}_{\Ebth{\balp}{\Ia}}
\,-\,
\,z_\mu\,\EBTH\cdot\bar{\DI}_{\EBTH}
\,\Ksim\,
2\,\Eth{\alp}{F}\,
\Ebth{\bbet}{D}\,\EsigD{\mu}{\alp}{\bbet}\,\Ebth{\dot{\gamma}}{F}\,\bar{\di}_{\Ebth{\dot{\gamma}}{D}}
\,,
\nn[0.75em]
\LIEVECsh{\ESSC{\alp}{\Ia}}
&=&
\,-\,
2\,\Eth{\bet}{\Ia}\,\ERsigU{\mu}{\nu}{\alp}{\bet}\,z_\mu\,\di_{z^{\nu}}
\,+\,
\Eth{\alp}{\Ia}\,\z\spr\di_{\z}
\,\Ksim\,
2\,\Eth{\alp}{F}\,\Eth{\bet}{\Ia}\,\Ebth{\dot{\gamma}}{F}\,\Esig{\mu}{\bet}{\dot{\gamma}}\,\di_{z^\mu}
\nonumber\\*[0.2em]
&&
\,-\,
z_\mu\,\EDsig{\mu}{\alp}{\bbet}\,\bar{\di}_{\Ebth{\bbet}{\Ia}}
\,+\,
2\,\Eth{\alp}{F}\,\Ebth{\bbet}{F}\,\bar{\di}_{\Ebth{\bbet}{\Ia}}
\,-\,
4\,\Eth{\alp}{F}\,\Eth{\bet}{\Ia}\di_{\Eth{\bet}{F}}
\,,
\nn[0.75em]
\LIEVECsh{\EBSS{\balp}{\Ia}}
&=&
\KSIM\,
2\,\Ebth{\bbet}{\Ia}\,\ERDsigU{\mu}{\nu}{\bbet}{\balp}\,z_\mu\,\di_{z^\nu}
\,\Ksim\,
\Ebth{\balp}{\Ia}\,\z\spr\di_{\z}
\,-\,
2\,\Ebth{\balp}{F}\,\Eth{\bet}{F}\,\Ebth{\dot{\gamma}}{\Ia}\,\Esig{\mu}{\bet}{\dot{\gamma}}\,\di_{z^\mu}
\nonumber\\*[0.2em]
& &
\,+\,
\,z_\mu\,\EDsig{\mu}{\bet}{\balp}\,\di_{\Eth{\bet}{\Ia}}
\,\Ksim\,
2\,\Ebth{\balp}{F}\,\Eth{\bet}{F}\,\di_{\Eth{\bet}{\Ia}}
\,\Ksip\,
4\,\Ebth{\balp}{F}\,\Ebth{\bbet}{\Ia}\,\bar{\di}_{\Ebth{\bbet}{F}}
\,,
\nn[0.75em]
\LIEVECsh{\RSYM^{\Ia}_{\Ib}}
&=&
\,-\,
\Eth{\alp}{\Ib}\,\di_{\Eth{\alp}{\Ia}}
\,+\,
\frac{1}{N}\delta^{\Ia}_{\Ib}\,\ETH\cdot\DI_{\ETH}
\,+\,
\Ebth{\balp}{\Ia}\,\bar{\di}_{\Ebth{\balp}{\Ib}}
\,-\,
\frac{1}{N}\delta^{\Ia}_{\Ib}\,\EBTH\cdot\bar{\DI}_{\EBTH}
\,,
\nn[0,75em]
\LIEVECsh{\RCHA}
&=&
\,\frac{1}{2}\ETH\cdot\DI_{\ETH}
\,-\,
\frac{1}{2}\EBTH\cdot\bar{\DI}_{\EBTH}
\,,
\nn[2em]
\LIECYCLsh{\ETR_\mu}
&=&
\LIECYCLsh{\EST{\alp}{\Ia}}
\,=\,
\LIECYCLsh{\EBST{\balp}{\Ia}}
\,=\,
0\,,
\quad
\LIECYCLsh{\EDL}
=
\EDL,\ \
\LIECYCLsh{\EROT{\mu}{\nu}}=\EROT{\mu}{\nu}
\,,
\nn[0.75em]
\LIECYCLsh{\ESC_{\mu}}
&=&
\ESC_{\mu}
\,-\,
2z_\mu\,\EDL
\,+\,
2\,z^\nu\,\EROT{\nu}{\mu}
\nn*[0.2em]
&&
\,\Ksip\,
\bigl(\,\Eth{\alp}{\Ia}\,\ERsigU{\nu}{\rho}{\bet}{\alp}\EsigD{\mu}{\bet}{\dot{\gamma}}\,\Ebth{\dot{\gamma}}{\Ia}
\,+\,
\Eth{\alp}{\Ia}\,\EsigD{\mu}{\alp}{\bbet}\ERDsigU{\nu}{\rho}{\dot{\gamma}}{\bbet}\,
\Ebth{\dot{\gamma}}{\Ia}\,\bigr)\,\EROT{\nu}{\rho}
\nn*[0.2em]
&&
\,-\,
\Eth{\alp}{\Ia}\,\EsigD{\mu}{\alp}{\bbet}\,\EBSS{\bbet}{\Ia}
\,+\,
\Ebth{\bbet}{\Ia}\,\EsigD{\mu}{\alp}{\bbet}\,\ESSC{\alp}{\Ia}
\nn*[0.2em]
&&
\,\Ksim\,
4\,\Eth{\alp}{F}\,\,\Ebth{\bbet}{D}\EsigD{\mu}{\alp}{\bbet}\,\RSYM^{F}_{D}
\,\Ksip\,
2\,(\frac{4}{N}-1)
\,\Eth{\alp}{\Ia}\,\Ebth{\bbet}{\Ia}\,\EsigD{\mu}{\alp}{\bbet}\,\RCHA
\,,
\nn[0.75em]
\LIECYCLsh{\ESSC{\alp}{\Ia}}
&=&
\ESSC{\alp}{\Ia}\,+\,
2\,\Eth{\alp}{\Ia}\,\EDL
\Ksim
2\,\Eth{\bet}{\Ia}\ERsigU{\mu}{\nu}{\alp}{\bet}\,\EROT{\mu}{\nu}
+
4\,\Eth{\alp}{F}\,\RSYM^{F}_{\Ia}
-
2\,\Bigl(\frac{4}{N}-1\Bigr)\,\Eth{\alp}{\Ia}\,\RCHA
\,,
\nn[0.75em]
\LIECYCLsh{\EBSS{\balp}{\Ia}}
&=&
\EBSS{\balp}{\Ia}
\Ksim
2\,\Ebth{\balp}{\Ia}\,\EDL
\Ksim 
2\,\Ebth{\bbet}{\Ia}\,\ERDsigU{\mu}{\nu}{\bbet}{\balp}\,\EROT{\mu}{\nu}
\Ksip
4\,\Ebth{\balp}{F}\,\RSYM^{\Ia}_{F}
\Ksim
2\,\Bigl(\frac{4}{N}-1\Bigr)\,\Ebth{\balp}{\Ia}\,\RCHA
\,,
\nn[0.75em]
\LIECYCLsh{\RSYM^{\Ia}_{\Ib}}
&=&
\RSYM^{\Ia}_{\Ib}\,,
\quad
\LIECYCLsh{\RCHA}
\,=\,
\RCHA\,.
\eeqa

%

\medskip

So we obtain an action of the superconformal Lie algebra on the super vertex operators.

Thus, we arrive to the following notion of a 
\textit{superconformal vertex algebra}.
It is a $(\Z/2\Z)$--graded vector space $V$ endowed with an even (bi)linear map 
$$
V \otimes V \, \ni \, a \otimes b \, \mapsto \, \EVO(a,\z,\ETH,\EBTH)\,b \, \in V [[\z]][(\z^2)^{-1}][\ETH,\EBTH]\,,
$$
an even vector $\vac \in V$ called a \textit{vacuum},
and an action on $V$ of the superconformal Lie algebra
keeping invariant (i.e., annihilating) the vacuum, such that
the coefficient fields in the expansion of $\EVO(a,\z,\ETH,\EBTH)$ in $\ETH$ and~$\EBTH$ fulfill all the axioms of vertex algebra of Sect. \ref{Se-VA},
and all the commutation relations (\ref{EqY-Z-com}) presenting the superconformal Lie algebra on the super vertex operators are satisfied.

\bigskip

\Section{Conclusion}

We developed an algebraic formalism for quantum superfields with extended superconformal symmetry analogous to vertex algebras.

This can have various applications.
First, in direction of cohomological analysis of anomalies in the perturbative models of such theories. Second, it gives a framework for constructing on shell models (i.e., models in a Hilbert space).

\bigskip
\noindent {\bf Acknowledgements.}
The author thank his adviser Prof. Nikolay Nikolov for useful directions and
discussions and Prof. Ivan Todorov for encouragement and useful discussions.

This work is partially supported by European Operational program HRD, contract BG051 PO001-3.3.06-0057.

\setcounter{section}{0}
\renewcommand{\thesection}{\Alph{section}}
\renewcommand{\theequation}{\Alph{section}.\arabic{equation}}
\renewcommand{\theTheorem}{\Alph{section}.\arabic{Theorem}}

\ASection{N-extended supreconformal Lie algebra}\label{superconformal_Lie_algebra_definiton}
\setcntrs
\ASubsection{Conformal Lie Algebra}\label{SeZ2.3}

Generators:

\begin{LIST}{33pt}
\item[$\bullet$] $\ETR_0,\dots,\ETR_{\DIM-1}$ -- generators of translations in the compact picture (cf. Eq. (\ref{TrInvAx})).
\item[$\bullet$] $\EROT{\mu}{\nu}$ ($0 \leqslant \mu < \nu \leqslant \DIM-1$) -- generators of rotations in the compact picture (cf. Eq. (\ref{CVA1})). 
\item[$\bullet$] $\EDL$ -- generator of dilatations in the compact picture (cf. Eq. (\ref{CVA2})). It is called \textit{conformal Hamiltonian}. The eigenvalues of the $\EDL$ are called scaling dimensions of the corresponding eigenstates (or fields).
\item[$\bullet$] $\ESC_0,\dots,\ESC_{\DIM-1}$ -- generators of special conformal transformations in the compact picture (cf. Eq. (\ref{CVA3})).
\end{LIST}

\medskip

\noindent Defining commutation relations:

\begin{gather}\label{ConCR}
\bigl[\EDL \hspace{1pt},\,\EDL\bigr]
\,=\,
\bigl[\EDL\hspace{1pt},\,\EROT{\mu}{\nu}\bigr]
\,=\,
\bigl[\ETR_{\mu}\hspace{1pt},\,\ETR_{\nu}\bigr]
\,=\
\bigl[\ESC_{\mu}\hspace{1pt},\,\ESC_{\nu}\bigr]
\,=\,0\,,
\nonumber\\[0.25em]
\bigl[\ETR_{\mu}\hspace{1pt},\,\ESC_{\nu}\bigr]
\,=\,
2\,\DEL{\mu}{\nu} \EDL - 2 \,\EROT{\mu}{\nu},
\\[0.25em]
\begin{array}{rclcrclc}
\bigl[\EROT{\mu}{\nu}\hspace{1pt},\,\ETR_{\gamma}\bigr]
\ravno
\DEL{\mu}{\gamma} \ETR_{\nu}\,-\,\DEL{\nu}{\gamma} \ETR_{\mu}\,,
&&
\bigl[\EROT{\mu}{\nu}\hspace{1pt},\,\ESC_{\gamma}\bigr]
\ravno
\DEL{\mu}{\gamma} \ESC_{\nu}\,-\,\DEL{\nu}{\gamma} \ESC_{\mu}\,,
&
\\[0.45em]
\bigl[\EDL\hspace{1pt},\, \ETR_{\mu}\bigr]
\ravno
\ETR_{\mu}\,,
&&
\bigl[\EDL\hspace{1pt},\,\ESC_{\mu}\bigr]
\ravno
-\,\ESC_{\mu}\,,
& 
\end{array}\nonumber
\\[0.25em]
\bigl[\EROT{\mu_1}{\nu_1}\hspace{1pt},\,\EROT{\mu_2}{\nu_2}\bigr]
\,=\,\DEL{\mu_1}{\mu_2}\hspace{1pt}\EROT{\nu_1}{\nu_2}
\,+\,
\DEL{\nu_1}{\nu_2}\hspace{1pt}\EROT{\mu_1}{\mu_2}
\,-\, \DEL{\mu_1}{\nu_2}\hspace{1pt}\EROT{\nu_1}{\mu_2}-\DEL{\nu_1}{\mu_2}\hspace{1pt}\EROT{\mu_1}{\nu_2}
\hspace{1pt},\nonumber
\end{gather}
where $\DEL{\mu}{\nu}$ is the diagonal Euclidean metric in $\R^{\DIM}$, i.e., the Kronecker delta symbol.

The real span of the generators (\ref{ConCR}) is the Euclidean conformal Lie algebra.
However, with respect to the ``physical'' real form the above generators are not real (cf. (\ref{conj-generators})) and so in our work we consider the complex linear span of them.

\ASubsection{Superconformal Lie Algebra}
This Lie superalgebra is extension of the conformal Lie algebra \eqref{ConCR}, written for $\DIM=4$, (it is in the even sector) with the following additional generators: 
\begin{LIST}{33pt}
\item[$\bullet$] Odd generators
$\EST{\alp}{\Ia}$ and $\EBST{\balp}{\Ia}$
called \textit{supertranslations}.($\alp=1,2$, $\balp=\dot{1},\dot{2}$, $\Ia=1,\ldots,N,$ as already explained in subsection \ref{Se2.1}.)
\item[$\bullet$] Odd generators
$\ESSC{\alp}{\Ia}$ and $\EBSS{\balp}{\Ia}$
called \textit{ super special conformal translations}.
The indices are as above.
\item[$\bullet$]
An even $U(1)$-generator $\RCHA$ called \textit{$R$--charge}.
\item[$\bullet$]
Even generators $\RSYM^{\Ia}_{\Ib}$ spanning the Lie algebra $\su(N)$ (i.e., $sl(N,\C)$, since we consider the complexified $\su(N)$). They are called \textit{$R$--sym\-m\-e\-t\-ry generators}.
\end{LIST}

Bellow we list the additional defining commutation relations split in few groups (Eqs. (\ref{ESCCR1}), (\ref{ESCCR2}), (\ref{ESCCR3}), (\ref{ESCCR4}), (\ref{ESCCR_5}) and (\ref{ESCCR6}) ).

\medskip
  
Polarization of the translations
(we remind that we use supercommutators):
\beq\label{ESCCR1}
\bigl[ \EST{\alp}{\Ia}\hspace{1pt},\, \EBST{\bbet}{\Ib}\bigr]
\, = \,
2 
\, \delta^{\Ia}_{\Ib} \, \Esig{\mu}{\alp}{\bbet} \, \ETR_\mu \,,
\eeq
where $\bigl\{\Esig{\mu}{\alp}{\bbet}\bigr\}$ is a $\Spin(4)$ intertwining operator
\beqa\label{Esig}
&
\bigl\{\Esig{\mu}{\alp}{\bbet} \bigr\} : 
\R^{4} \, \to \, \Hom\bigl(\ESPIN,\EBSPIN)\bigr)\, (\, \cong \, \ESPIN^* \otimes \EBSPIN) 
\,, & \quad
\bnn
&
\EsigD{\mu}{\alp}{\bbet} \, := \,
\delta_{\mu,\mu'} \, \Esig{\mu'}{\alp}{\bbet} \, \equiv \, \Esig{\mu}{\alp}{\bbet}
\,. & \quad
\eeqa
Here $\ESPIN$ and $\EBSPIN$ are, respectively, (equivalent to) $(\frac{1}{2},0)$ and $(0,\frac{1}{2})$  representations of the orthogonal Lie algebra:
\beq\label{ESPIN-act}
\bigl\{\ERsig{\mu}{\nu}{\alp}{\bet}\bigr\} \, : \, \ESPIN \, \mathop{\longrightarrow}\, \ESPIN \,,
\eeq
\begin{gather}
\ERsig{\mu_1}{\nu_1}{\alp}{\gamma}\,\ERsig{\mu_2}{\nu_2}{\gamma}{\bet}
\,-\,
\ERsig{\mu_2}{\nu_2}{\alp}{\gamma}\,\ERsig{\mu_1}{\nu_1}{\gamma}{\bet}
\,=\,\label{ESPIN-representation}\\
\,+\,
\delta_{\mu_1,\mu_2}\,\ERsig{\nu_1}{\nu_2}{\alp}{\bet}
\,+\,
\delta_{\nu_1,\nu_2}\,\ERsig{\mu_1}{\mu_2}{\alp}{\bet}
\,-\,
\delta_{\mu_1,\nu_2}\,\ERsig{\nu_1}{\mu_2}{\alp}{\bet}
\,-\,
\delta_{\nu_1,\mu_2}\,\ERsig{\mu_1}{\nu_2}{\alp}{\bet}\nonumber
\end{gather}
and
\beq\label{ESPIN-c-act}
\bigl\{\ERDsig{\mu}{\nu}{\balp}{\bbet}\bigr\} \, : \, \EBSPIN \, \mathop{\longrightarrow}
\, \EBSPIN \,,
\eeq
\begin{gather}
\ERDsig{\mu_1}{\nu_1}{\balp}{\dot{\gamma}}\,\ERDsig{\mu_2}{\nu_2}{\dot{\gamma}}{\bbet}
\,-\,
\ERDsig{\mu_2}{\nu_2}{\balp}{\dot{\gamma}}\,\ERDsig{\mu_1}{\nu_1}{\dot{\gamma}}{\bbet}
\,=\,\label{EBSPIN-representation}\\
\,+\,
\delta_{\mu_1,\mu_2}\,\ERDsig{\nu_1}{\nu_2}{\balp}{\bbet}
\,+\,
\delta_{\nu_1,\nu_2}\,\ERDsig{\mu_1}{\mu_2}{\balp}{\bbet}
\,-\,
\delta_{\mu_1,\nu_2}\,\,\ERDsig{\nu_1}{\mu_2}{\balp}{\bbet}
\,-\,
\delta_{\nu_1,\mu_2}\,\ERDsig{\mu_1}{\nu_2}{\balp}{\bbet}\,.\nonumber
\end{gather}
The intertwining property of $\bigl\{\Esig{\mu}{\alp}{\bbet}\bigr\}$ reads
\beq\label{intertw}
\EsigD{\mu}{\alp}{\bbet}\,\delta_{\nu,\rho}-\EsigD{\nu}{\alp}{\bbet}\,\delta_{\mu,\rho}
\,=\,
\EsigD{\rho}{\gamma}{\bbet}\,\ERsig{\mu}{\nu}{\gamma}{\alp}
-
\ERDsig{\mu}{\nu}{\bbet}{\dot{\sigma}}\,\EsigD{\rho}{\alp}{\dot{\sigma}}\,.
\eeq
The $\Est-\Est$--relations are completed by
\beq\label{ESCCR2}
\bigl[ \EST{\alp}{\Ia}\hspace{1pt},\, \EST{\bet}{\Ib}\bigr] \, = \, 0 \, = \,
\bigl[ \EBST{\balp}{\Ia}\hspace{1pt},\, \EBST{\bbet}{\Ib}\bigr] \,.
\eeq
We note that in the so called super-Poincar\'e algebra it is admitted to have nonzero relations instead of (\ref{ESCCR2}), which generate the $R$--symmetry.
However, in the superconformal algebra 
it is not the case
and the $R$--symmetry appears in the $\Est-\Essc$--relations instead (cf. Eq. (\ref{ESCCR6}) below).

The relations between the $\Est$'s and the even generators are
\beq\label{ESCCR3}
\hspace{0pt}
\begin{array}{rclcrclc}
\bigl[ \ETR_{\mu} \hspace{1pt},\, \EST{\alp}{\Ia} \bigr]
\ravno
0
\,, &&
\bigl[ \ETR_{\mu} \hspace{1pt},\, \EBST{\balp}{\Ia} \bigr]
\ravno
0
\,,& \\
\bigl[ \EROT{\mu}{\nu} \hspace{1pt},\, \EST{\alp}{\Ia} \bigr]
\ravno
\ERsig{\mu}{\nu}{\bet}{\alp} \ \EST{\bet}{\Ia}
\,, &&
\bigl[ \EROT{\mu}{\nu} \hspace{1pt},\, \EBST{\balp}{\Ia} \bigr]
\ravno
-
\ERDsig{\mu}{\nu}{\balp}{\bbet} \ \EBST{\bbet}{\Ia}
\,,& \raisebox{12pt}{}\\
\bigl[ \EDL \hspace{1pt},\, \EST{\alp}{\Ia} \bigr]
\ravno
\frac{1}{2} \, \EST{\alp}{\Ia}
\,, &&
\bigl[ \EDL \hspace{1pt},\, \EBST{\balp}{\Ia} \bigr]
\ravno
\frac{1}{2} \, \EBST{\balp}{\Ia}
\,,& \raisebox{12pt}{}\\
\bigl[ \ESC_{\mu} \hspace{1pt},\, \EST{\alp}{\Ia} \bigr]
\ravno
- 
\EsigD{\mu}{\alp}{\bbet} \, \EBSS{\bbet}{\Ia}
\,, &&
\bigl[ \ESC_{\mu} \hspace{1pt},\, \EBST{\balp}{\Ia} \bigr]
\ravno
\EsigD{\mu}{\bet}{\balp} \, \ESSC{\bet}{\Ia}
\,,& \raisebox{12pt}{}\\
\bigl[ \RCHA \hspace{1pt},\, \EST{\alp}{\Ia} \bigr]
\ravno
\frac{1}{2} \, \EST{\alp}{\Ia}
\,, &&
\bigl[ \RCHA \hspace{1pt},\, \EBST{\balp}{\Ia} \bigr]
\ravno
-\frac{1}{2} \, \EBST{\balp}{\Ia}
\,,& \raisebox{12pt}{}\\
\bigl[ \RSYM_{\Ic}^{\Ib} \hspace{1pt},\, \EST{\alp}{\Ia} \bigr]
\ravno
-(\delta^{\Ia}_{\Ic} \, \EST{\alp}{\Ib}-\frac{1}{N}\delta^{\Ib}_{\Ic}\,\EST{\alp}{\Ia})
\,, &&
\bigl[ \RSYM_{\Ic}^{\Ib} \hspace{1pt},\, \EBST{\balp}{\Ia} \bigr]
\ravno
(\delta^{\Ib}_{\Ia} \, \EBST{\balp}{\Ic}-\frac{1}{N}\delta^{\Ib}_{\Ic}\,\EBST{\balp}{\Ia})
\,,&\raisebox{12pt}{}\\[0.3em]
\end{array}
\eeq
where $\ERsig{\mu}{\nu}{\alp}{\bet}$ and $\ERDsig{\mu}{\nu}{\balp}{\bbet}$ are defined in \eqref{ESPIN-act} and \eqref{ESPIN-c-act}.

As we mentioned above, the additional even generators $\RSYM^{\Ia}_{\Ib}$ are $\su(N)$ generators, i.e.,
\beq\label{RSYMgen}
\bigl[ \RSYM^{\Ia}_{\Ib} \hspace{1pt},\, \RSYM^{\Ia_1}_{\Ib_1} \bigr]
\, = \,
\delta^{\Ia}_{\Ib_1} \, \RSYM^{\Ia_1}_{\Ib}
\, - \,
\delta^{\Ia_1}_{\Ib} \, \RSYM^{\Ia}_{\Ib_1}
\eeq
and $\mathop{\sum}\limits_{\Ia=1}^N\RSYM^{\Ia}_{\Ia}=0$.
The $R$-charge $\RCHA$ is central element in the even part, and also,
the $R$--symmet\-ry generators commute with the conformal Lie algebra generators,
\beqa\label{ESCCR4}
\podr
\bigl[ \RCHA \hspace{1pt},\, \RSYM^{\Ia}_{\Ib} \bigr]
\, = \,
\bigl[ \RCHA \hspace{1pt},\, \ETR_{\mu} \bigr]
\, = \,
\bigl[ \RCHA \hspace{1pt},\, \EROT{\mu}{\nu} \bigr]
\, = \,
\bigl[ \RCHA \hspace{1pt},\, \EDL \bigr]
\, = \,
\bigl[ \RCHA \hspace{1pt},\, \ESC_{\mu} \bigr]
\nnb
\podr = \,
\bigl[ \RSYM^{\Ia}_{\Ib} \hspace{1pt},\, \ETR_{\mu} \bigr]
\, = \,
\bigl[ \RSYM^{\Ia}_{\Ib} \hspace{1pt},\, \EROT{\mu}{\nu} \bigr]
\, = \,
\bigl[ \RSYM^{\Ia}_{\Ib} \hspace{1pt},\, \EDL \bigr]
\, = \,
\bigl[ \RSYM^{\Ia}_{\Ib} \hspace{1pt},\, \ESC_{\mu} \bigr]
\, = \, 0 \,.
\eeqa

Next, the commutation relations of the other odd generators $\ESSC{\alp}{\Ia}$ and $\EBSS{\balp}{\Ia}$
have a certain similarity with those for the $\Est$'s:
\begin{gather}
\bigl[ \ESSC{\alp}{\Ia}\hspace{1pt},\, \EBSS{\bbet}{\Ib}\bigr]
\, = \,
2 
\, \delta^{\Ib}_{\Ia} \, \EDsig{\mu}{\alp}{\bbet} \, \ESC_\mu \,,
\medskip\nnb
\bigl[ \ESSC{\alp}{\Ia}\hspace{1pt},\, \ESSC{\bet}{\Ib}\bigr] \, = \, 0 \, = \,
\bigl[ \EBSS{\balp}{\Ia}\hspace{1pt},\, \EBSS{\bbet}{\Ib}\bigr] \,,
\medskip\nnb
\begin{array}{rclcrclc}
\bigl[ \ETR_{\mu} \hspace{1pt},\, \ESSC{\alp}{\Ia} \bigr]
\ravno
\EDsigD{\mu}{\alp}{\bbet} \, \EBST{\bbet}{\Ia}
\,, &&\ \
\bigl[ \ETR_{\mu} \hspace{1pt},\, \EBSS{\balp}{\Ia} \bigr]
\ravno
- 
\EDsigD{\mu}{\bet}{\balp} \, \EST{\bet}{\Ia}
\,,& \\
\bigl[ \EROT{\mu}{\nu} \hspace{1pt},\, \ESSC{\alp}{\Ia} \bigr]
\ravno
-
\ERsig{\mu}{\nu}{\alp}{\bet} \ \ESSC{\bet}{\Ia}
\,, &&\ \
\bigl[ \EROT{\mu}{\nu} \hspace{1pt},\, \EBSS{\balp}{\Ia} \bigr]
\ravno
\ERDsig{\mu}{\nu}{\bbet}{\balp} \ \EBSS{\bbet}{\Ia}
\,,& \raisebox{12pt}{}\\
\bigl[ \EDL \hspace{1pt},\, \ESSC{\alp}{\Ia} \bigr]
\ravno
-\frac{1}{2} \, \ESSC{\alp}{\Ia}
\,, &&\ \
\bigl[ \EDL \hspace{1pt},\, \EBSS{\balp}{\Ia} \bigr]
\ravno
-\frac{1}{2} \, \EBSS{\balp}{\Ia}
\,,& \raisebox{12pt}{}\\
\bigl[ \ESC_{\mu} \hspace{1pt},\, \ESSC{\alp}{\Ia} \bigr]
\ravno
0
\,, &&\ \
\bigl[ \ESC_{\mu} \hspace{1pt},\, \EBSS{\balp}{\Ia} \bigr]
\ravno
0
\,,& \raisebox{12pt}{}\\
\bigl[ \RCHA \hspace{1pt},\, \ESSC{\alp}{\Ia} \bigr]
\ravno
- \frac{1}{2} \, \ESSC{\alp}{\Ia}
\,, &&\ \
\bigl[ \RCHA \hspace{1pt},\, \EBSS{\balp}{\Ia} \bigr]
\ravno
\frac{1}{2} \, \EBSS{\balp}{\Ia}
\,,& \raisebox{12pt}{}\\
\bigl[ \RSYM_{\Ic}^{\Ib} \hspace{1pt},\, \ESSC{\alp}{\Ia} \bigr]
\ravno
(\delta^{\Ib}_{\Ia} \, \ESSC{\alp}{\Ic}-\frac{1}{N}\delta^{\Ib}_{\Ic}\, \ESSC{\alp}{\Ia})
\,, &&\ \
\bigl[ \RSYM_{\Ic}^{\Ib} \hspace{1pt},\, \EBSS{\balp}{\Ia} \bigr]
\ravno
-(\delta_{\Ic}^{\Ia} \, \EBSS{\balp}{\Ib}-\frac{1}{N}\delta^{\Ib}_{\Ic}\, \EBSS{\balp}{\Ia})
\,.&\raisebox{12pt}{}\\[0.3em]
\end{array}
\hspace{-5pt}\label{ESCCR_5}
\end{gather}
Here, $\{\EDsig{\mu}{\alp}{\bbet}\}$ is an intertwining operator:
\beqa
& 
\bigl\{\EDsig{\mu}{\alp}{\bbet} \bigr\} : 
\R^{4} \, \to \, \Hom\bigl(\EBSPIN,\ESPIN)\bigr)\, (\, \cong \, \ESPIN \otimes \EBSPIN^{\ *}) 
\,, &
\bnn
&
\EDsigD{\mu}{\alp}{\bbet} \, := \,
\delta_{\mu,\mu'} \, \EDsig{\mu'}{\alp}{\bbet} \, \equiv \, \EDsig{\mu}{\alp}{\bbet}
\,, &
\label{EsigD}\\[0.3em]
&
\EDsigD{\mu}{\alp}{\bbet}\,\delta_{\nu,\rho}-\EDsigD{\nu}{\alp}{\bbet}\,\delta_{\mu,\rho}
\,=\,
-
\ERsig{\mu}{\nu}{\alp}{\gamma}\,\EDsigD{\rho}{\gamma}{\bbet}
+
\EDsigD{\rho}{\alp}{\dot{\sigma}}\,\ERDsig{\mu}{\nu}{\dot{\sigma}}{\bbet}\,.
&
\quad\label{intertwD}
\eeqa

The remaining $\Est-\Essc$ relations are:
\beqa\label{ESCCR6}
\bigl[ \EST{\alp}{\Ia} \hspace{1pt},\, \ESSC{\bet}{\Ib} \bigr]\,
\ravno \,
-2\,
\delta^{\Ia}_{\Ib}\,
\bigl(\delta_{\alp}^{\bet}\,\EDL
-
\ERsigU{\mu}{\nu}{\bet}{\alp} \, \EROT{\mu}{\nu}\bigr)
-4\,\delta_{\alp}^{\bet}
\RSYM_{\Ib}^{\Ia}
+ 2\hspace{1pt}\Bigl( \frac{4}{N}-1\Bigr)\hspace{1pt}
\delta^{\Ia}_{\Ib}\,\delta_{\alp}^{\bet}
\, \RCHA\,,
\nnb
\bigl[ \EBST{\balp}{\Ia} \hspace{1pt},\, \EBSS{\bbet}{\Ib} \bigr]\,
\ravno \,
2\,\delta_{\Ia}^{\Ib}\,
\bigl(\delta_{\bbet}^{\balp} \, \EDL
+
\ERDsigU{\mu}{\nu}{\balp}{\bbet} \, \EROT{\mu}{\nu}{\big)}
-4\,\delta_{\bbet}^{\balp}
\RSYM_{\Ia}^{\Ib}
+ 2\hspace{1pt}\Bigl( \frac{4}{N}-1\Bigr)\hspace{1pt} 
\delta_{\Ia}^{\Ib}\,\delta_{\bbet}^{\balp}
\, \RCHA\,,
\nnb
&&\bigl[ \EST{\alp}{\Ia} \hspace{1pt},\, \EBSS{\bbet}{\Ib} \bigr]
= \, 0 \, = \,
\bigl[ \EBST{\balp}{\Ia} \hspace{1pt},\, \ESSC{\bbet}{\Ib} \bigr]
\,.
\eeqa

\medskip

\begin{Proposition}\label{Pr3N1}
Relations \eqref{ConCR}, \eqref{ESCCR1}, \eqref{ESCCR2}--\eqref{ESCCR_5} and \eqref{ESCCR6} define a Lie superalgebra structure on \[\Span_\C\{\ETR_{\mu},\ESC_{\mu},\EROT{\mu}{\nu},\EDL,\EST{\alp}{\Ia},\EBST{\balp}{\Ia},\ESSC{\alp}{\Ia},\EBSS{\balp}{\Ia},\RCHA,\RSYM^{\Ia}_{\Ib}\,\,\vline\,\, \forall\,
\text{admissible } \mu,\nu,\alp,\balp,\Ia,\Ib\}\] if (and only if) the following relations are satisfied for the structure constants $\Esig{\mu}{\alp}{\bbet}$, $\EDsig{\mu}{\alp}{\bbet}$, $\ERsig{\mu}{\nu}{\alp}{\bet}$ and $\ERDsig{\mu}{\nu}{\balp}{\bbet}$:
\beqa
\EDsigD{\mu}{\gamma}{\bbet}\EsigD{\nu}{\alp}{\bbet}+
\EDsigD{\nu}{\gamma}{\bbet}\EsigD{\mu}{\alp}{\bbet}
&=&
2\,\delta_{\mu,\nu}\,\delta^{\gamma}_{\alp}
\label{anitcommute_sigma1}
\,,
\\
\ERsig{\mu}{\nu}{\gamma}{\alpha}&=&
-\,\frac{1}{4}\bigl(\,\EDsigD{\mu}{\gamma}{\bbet}\EsigD{\nu}{\alp}{\bbet}        
- \EDsigD{\nu}{\gamma}{\bbet}\EsigD{\mu}{\alp}{\bbet}\,\bigr)\label{expression_sigma_mu_nu} 
\,,
\\
\EsigD{\mu}{\bet}{\balp}\EDsigD{\nu}{\bet}{\dot{\gamma}}+
\EsigD{\nu}{\bet}{\balp}\EDsigD{\mu}{\bet}{\dot{\gamma}}
&=&
2\,\delta_{\mu,\nu}\,\delta^{\balp}_{\dot{\gamma}}
\label{anticommute_sigma2}
\,,
\\
\ERDsig{\mu}{\nu}{\balp}{\dot{\gamma}}
&=&
-\,\frac{1}{4}\bigl(\,\EsigD{\mu}{\bet}{\balp}\EDsigD{\nu}{\bet}{\dot{\gamma}}
-\EsigD{\nu}{\bet}{\balp}\EDsigD{\mu}{\bet}{\dot{\gamma}}\,\bigr)\label{expression_barsigma_mu_nu}
\,,
\\
\EDsig{\mu}{\bet}{\dot{\gamma}}\EsigD{\mu}{\alp}{\dot{\sigma}}&=&
2\delta^{\bet}_{\alp}\delta^{\dot{\sigma}}_{\dot{\gamma}}
\,.
\label{orthogonality_of_sigma_barsigma}
\eeqa
In fact, Eqs. \eqref{anitcommute_sigma1} and \eqref{anticommute_sigma2} just define a Clifford algebra structure on $\ESPIN\oplus\EBSPIN$ by setting
\beq\label{GAMMA-MATR}
\Gamma_{\mu} := 
\left(
\begin{array}{cc}
0 & \EDsig{\mu}{}{}
\\
\Esig{\mu}{}{} & 0
\end{array}
\right)
: \ESPIN\oplus\EBSPIN \longrightarrow \ESPIN\oplus\EBSPIN
\,.
\eeq
\end{Proposition}

\bigskip

\noindent
\textit{Sketch of the proof.}
The (super) Jacobi identities for $\bigl[\ETR_{\mu},\,[\ESC_{\nu},\,\EST{\alp}{\Ia}]\,\bigr]$ and
$\bigl[\ESC_{\mu},\,[\ETR_{\nu},\,\ESSC{\alp}{\Ia}]\,\bigr] $ are equivalent to \eqref{anitcommute_sigma1} and \eqref{expression_sigma_mu_nu}. The (super) Jacobi identities for
$\bigl[\ETR_{\mu},\,[\ESC_{\nu},\,\EBST{\balp}{\Ia}]\,\bigr]$ and
$\bigl[\ESC_{\mu},\,[\ETR_{\nu},\,\EBSS{\alp}{\Ia}]\,\bigr] $ are  equivalent to \eqref{anticommute_sigma2} and \eqref{expression_barsigma_mu_nu}. Finally, the (super) Jacobi identity for $\bigl[\,\EST{\alp}{\Ia},\,[\ESSC{\bet}{B},\EBSS{\dot{\gamma}}{C}]\,\bigr]$ (using also \eqref{anticommute_sigma2}) is equivalent to \eqref{orthogonality_of_sigma_barsigma}.$\quad\square$
\par We note that \eqref{expression_sigma_mu_nu}, \eqref{anticommute_sigma2},   \eqref{expression_barsigma_mu_nu} implicate relations \eqref{intertw}, \eqref{intertwD}, \eqref{ESPIN-representation}. In our calculations we make an extensive use the following relations 
\begin{gather}\label{eq3.22-n}
\begin{array}{rcl}
\ERsigU{\mu}{\nu}{\bet}{\alp}\ERDsig{\mu}{\nu}{\dot{\sigma}}{\dot{\gamma}}
\ravno
0,
\\[0.2em]
\ERDsigU{\mu}{\nu}{\bbet}{\balp}\ERDsig{\mu}{\nu}{\dot{\gamma}}{\dot{\sigma}}
\ravno
\delta^{\bbet}_{\balp}\,\delta^{\dot{\gamma}}_{\dot{\sigma}}
-2\delta^{\dot{\gamma}}_{\balp}\,\delta^{\bbet}_{\dot{\sigma}},
\\[0.2em]
\ERsigU{\mu}{\nu}{\bet}{\alp}\ERsig{\mu}{\nu}{\sigma}{\gamma}
\ravno
\delta^{\bet}_{\alp}\,\delta^\sigma_\gamma
-2\delta^{\bet}_{\gamma}\,\delta^{\sigma}_{\alp}\,.
\end{array}\\
\EsigD{\mu}{\alp}{\bbet}\EsigD{\nu}{\gamma}{\dot{\sigma}}
\,+\,
\EsigD{\nu}{\alp}{\bbet}\EsigD{\mu}{\gamma}{\dot{\sigma}}
\,-\,
\EsigD{\mu}{\alp}{\dot{\sigma}}\EsigD{\nu}{\gamma}{\bbet}
\,-\,
\EsigD{\nu}{\alp}{\dot{\sigma}}\EsigD{\mu}{\gamma}{\bbet}
\,=\,
\EsigD{\rho}{\alp}{\bbet}\Esig{\rho}{\gamma}{\dot{\sigma}}\DEL{\mu}{\nu},
\nonumber
\end{gather}
which are corollaries of Eqs. \eqref{expression_sigma_mu_nu} -- \eqref{orthogonality_of_sigma_barsigma}.

\medskip

Realization of  the matrices $\Esig{\mu}{}{}$ and $\EDsig{\mu}{}{},\quad\! \mu=0,1,2,3$:%
\begin{gather}\label{sigma_realization}
\begin{array}{lcl}
\Esig{0}{}{}\,=\,\begin{pmatrix}
                  \phantom{-}0&-1\\
                  \phantom{-}1&\phantom{-}0
                  \end{pmatrix},
&\qquad & \Esig{1}{}{}\,=\,\begin{pmatrix}
                  -i&\phantom{-}0\\
                  \phantom{-}0&\phantom{0}i
                  \end{pmatrix},\\[1.5em]
\Esig{2}{}{}\,=\,\begin{pmatrix}
                  -1&\phantom{-}0\\
                  \phantom{-}0&-1
                  \end{pmatrix},
&\quad & \Esig{3}{}{}\,=\,\begin{pmatrix}
                  \phantom{-}0&\phantom{-}i\\
                  \phantom{-}i&\phantom{-}0
                  \end{pmatrix},
\end{array}
\end{gather} 

$ $
\begin{gather}\label{sigma_bar_realization}
\begin{array}{lcl}
\EDsig{0}{}{}\,=\,\begin{pmatrix}
                  \phantom{-}0&\phantom{-}1\\
                   -1&\phantom{-}0
                  \end{pmatrix},
&\qquad & \EDsig{1}{}{}\,=\,\begin{pmatrix}
                  \phantom{-}i&\phantom{-}0\\
                  \phantom{-}0&-i
                  \end{pmatrix},\\[1.5em]
\EDsig{2}{}{}\,=\,\begin{pmatrix}
                  -1&\phantom{-}0\\
                  \phantom{-}0&-1
                  \end{pmatrix},
&\quad & \EDsig{3}{}{}\,=\,\begin{pmatrix}
                  \phantom{-}0&-i\\
                  -i&\phantom{-}0
                  \end{pmatrix}.
\end{array}
\end{gather}

This matrices are related to the Pauli matrices in the following way:
\beq
\Esig{\mu}{\bet}{\balp}=
(\bar{\sigma}^{\mu}_{E})^{\balp\hspace{1pt}\gamma}\varepsilon_{\gamma\hspace{1pt}\bet}\hspace{1pt},
\qquad
\EDsig{\mu}{\alp}{\bbet}=
\varepsilon^{\alp\hspace{1pt}\gamma}(\sigma^{\mu}_{E})_{\gamma\hspace{1pt}\bbet},
\nonumber
\eeq
where $\varepsilon_{\alp\hspace{1pt}\gamma}$  $(\varepsilon^{12}=-\varepsilon^{21}=-\varepsilon_{12}=\varepsilon_{21}=1)$ is spinor metric tensor,
\beq
\sigma_E=(\mathds{1},i\sigma^1, i\sigma^2, i\sigma^3),
\qquad
\bar{\sigma}_E=(\mathds{1},-i\sigma^1,-i\sigma^2,-i\sigma^3)
\nonumber
\eeq
and $\sigma^1,\sigma^2,\sigma^3$ are the Pauli matrices:
\beq
\begin{array}{ccc}
\sigma^1\,=\,\begin{pmatrix}
             \phantom{-}0 &\phantom{-}1\\
             \phantom{-}1 &\phantom{-}0
             \end{pmatrix},&
\sigma^2\,=\,\begin{pmatrix}
             \phantom{-}0 &-i\\
             \phantom{-}i &\phantom{-}0
             \end{pmatrix},&
\sigma^3\,=\,\begin{pmatrix}
             \phantom{-}1 &\phantom{-}0\\
             \phantom{-}0 &-1
             \end{pmatrix}. 
\end{array}\nonumber
\eeq 
\ASubsection{Real structure}\label{SeZ4}
As we mentioned in the introduction to Sect. \ref{Se2}, the relevant real structure in the compact picture is not an ordinary complex conjugation. It is given by:

\beq \label{conj-generators}
\ETR_\mu^\star =\ESC_\mu\,,
\quad
\EDL^\star =-\EDL\,,
\quad
\EROT{\mu}{\nu}^\star =\EROT{\mu}{\nu\,},
\qquad
\eeq

\begin{gather}\label{real_atructure_odd_definite}
\begin{array}{rclcrclc}
(\RSYM^{\Ib}_{\Ic})^\star\ravno -\RSYM^{\Ic}_{\Ib}\,, &&
\RCHA^\star\ravno-\RCHA\,, &
\\[0.2em]
(\EST{\alp}{\Ia})^\star\ravno i\,\ESSC{\alp}{\Ia}\,, &&
(\EBST{\balp}{\Ia})^\star\ravno -i\,\EBSS{\balp}{\Ia}\,, &
\\[0.2em]
(\ESSC{\alp}{\Ia})^\star\ravno i\,\EST{\alp}{\Ia}\,, & &
(\EBSS{\balp}{\Ia})^\star\ravno -i\,\EBST{\balp}{\Ia}\,, &
\end{array}
\end{gather}

where $\star$ is an antilinear involution
\beqa
&
A^{\star\star} \!= A\,,
\!\quad\!
[A_1,A_2]^\star \!=[A_1^\star,A_2^\star]\,,
\!\quad\!
(\alpha_1 A_1+\alpha_2 A_2)^\star \!=\bar{\alpha}_1A_1^\star+\bar{\alpha}_2A_2^\star\,,
&\label{conj-alg}
\qquad
\eeqa
$A$, $A_1$, $A_2$ - arbitrary elements of the suprconformal Lie algebra and $\alpha_1$, $\alpha_2$ -  complex numbers.

It imposes additional relations between structure constants:

\beqa \label{sigma_sigma_bar_relations}
\ERDsig{\mu}{\nu}{\balp}{\bbet}
\ravno
-\overline{\ERDsig{\mu}{\nu}{\bbet}{\balp}}\,,
\nn
\ERsig{\mu}{\nu}{\alp}{\bet}
\ravno
-\overline{\ERsig{\mu}{\nu}{\bet}{\alp}}\,,
\nn
\EDsig{\mu}{\alp}{\bbet}
\ravno
\overline{\Esig{\mu}{\alp}{\bbet}}\,,
\eeqa
which are fulfilled by our realization \eqref{sigma_realization}, \eqref{sigma_bar_realization} of the $\Sigma$ - matrices.

The conjugation of the coordinates is given by:
\begin{eqnarray}
\bigl(z_{\pm}^\mu \bigr)^*\,
&=&
\frac{z_{\mp}^\mu}{(\z_{\mp})^2}\,,
\\[0.2em]
\bigl(\Eth{\alp}{\Ia}\bigr)^*
&=&
i\frac{z_{-}^\mu}{(\z_{-})^2}\EsigD{\mu}{\alp}{\bbet}\Ebth{\bbet}{\Ia}\,,
\\[0.2em]
\bigl(\Ebth{\balp}{\Ia}\bigr)^*
&=&
i\frac{z_{+}^\mu}{(\z_{+})^2}\EsigD{\mu}{\bet}{\balp}\Eth{\bet}{\Ia}\,,
\end{eqnarray}
where
\beq
z_{\pm}^\mu\,:=\,z^\mu\,\pm\,\Eth{\alp}{\Ia}\Esig{\mu}{\alp}{\bbet}\Ebth{\bbet}{\Ia}.
\eeq

\end{document}